\documentclass[aps,showkeys,superscriptaddress,preprint,tightenlines]{revtex4}
\usepackage{times}
\usepackage{amsmath}
\usepackage{graphicx}
\usepackage{epsfig}
\usepackage{dcolumn}
\usepackage{bm}
\usepackage{color}
\usepackage{longtable}
\usepackage{array}
\usepackage{multirow}

\newcommand{\BR}{{\cal B}}

\newcommand{\pbar}{\bar{p}}

\newcommand{\nbar}{\bar{n}}

\newcommand{\nbp}{\bar{n}p}
\newcommand{\nbn}{\bar{n}n}

\newcommand{\jpsi}{J\kern-0.15em/\kern-0.15em\psi\kern0.15em}
\newcommand{\psp}{\psi(2S)}

\newcommand{\EE}{e^+e^-}

\newcommand{\ppinb}{p\pi^-\bar{n}}

\newcommand{\beq}{\begin{equation}}
\newcommand{\eeq}{\end{equation}}
\newcommand{\bitm}{\begin{itemize}}
\newcommand{\eitm}{\end{itemize}}

\begin{document}

\title{\boldmath
A cornucopia of antineutrons and hyperons from super $\jpsi$ factory \\
for next-generation nuclear and particle physics high-precision experiments
}
\author{Chang-Zheng Yuan}
 \email{yuancz@ihep.ac.cn}
 \affiliation{Institute of High Energy Physics, Chinese Academy of Sciences,
 Beijing 100049, China}
 \affiliation{University of Chinese Academy of Sciences, Beijing 100049, China}
\author{Marek Karliner}
\email{marek@tauex.tau.ac.il}
\affiliation{School of Physics and Astronomy, Tel Aviv University, Tel Aviv 69978, Israel}

\begin{abstract}

In order to study the interactions and structure of various types
of matter one typically needs to carry out scattering experiments
utilizing many different particles as projectiles. Whereas beams
of $e^\pm$, $\mu^\pm$, $\pi^\pm$, $K^\pm$, proton, antiproton, and
of various heavy ions have been produced and have enabled many
scientific breakthroughs, beams of antineutrons, hyperons
($\Lambda$, $\Sigma$ and $\Xi$) and their antiparticles are
typically not easy to obtain. Here we point out and investigate a
new high-quality source of these particles: a super $\jpsi$
factory with capability of accumulating trillions of $\jpsi$
decays each year. In the relevant $\jpsi$ decays the desired
particle is produced together with other final state particles
that can be tagged. This allows accurate determination of the flux
and momentum of the projectile, enabling unprecedented
precision-study of the corresponding interactions with a broad
range of targets. These novel high-statistics sources of baryons
and antibaryons with precisely known kinematics open fresh
opportunities for applications in particle and nuclear physics,
including antinucleon-nucleon interaction, nonvalence $s\bar{s}$
component of the nucleon, (anti)hyperon-nucleon interaction, OZI
violation, (multi-strange) hypernuclei, exotic light hadron
spectroscopy and many others, as well as calibration of Monte
Carlo simulation for hadronic and medical physics.

\end{abstract}


\date{\today}

\maketitle


Scattering experiments using many different kinds of beams are the
mainstay of experiments investigating fundamental interactions and
structure of matter at subatomic level. Beams of long-lived
charged particles and of photons are easy to produce and so many
experiments using charged projectiles have been carried out during
more than 100 years since the trailblazing experiment shooting
$\alpha$ particles into gold foil enabled Rutherford to infer the
existence of the atomic nucleus~\cite{rutherford}. Since then,
$e^\pm$, $\mu^\pm$, $\pi^\pm$, $K^\pm$, proton, antiproton, photon
and various heavy ion beams have been produced and have served as
enablers of many scientific breakthroughs. Beams of some neutral
particles, like neutrons and neutrinos are relatively easy to
produce, but difficult to control, i.e. have large momentum
spread. Beams of other neutral particles, such as antineutron and
$K^0/\bar{K}^0$, and of long-lived hyperons \,($\Lambda$,
$\Sigma^{+,-}$, $\Xi^{0,-}$)\, and their antiparticles
\,($\bar\Lambda$, $\bar{\Sigma}^{-,+}$, $\bar{\Xi}^{0,+}$)\, have
great physics potential, but typically are much more difficult to
produce and control.

Although potentially extremely useful for investigating
nonperturbative QCD and nuclear structure, experimental studies
using antineutron beams have been very limited till now, due to
the severe difficulties in accumulating sufficient number of
antineutrons with known flux and momentum~\cite{Bressani:2003pv}.
The best antineutron sources so far have been obtained at BNL
E-767~\cite{E767} (with momentum between 100 and 500~MeV/$c$) and
CERN OBELIX experiments~\cite{OBELIX} (with momentum between 50
and 400~MeV/$c$), enabling quite a wide range of physics topics to
be studied, from nuclear physics to hadron spectroscopy,  albeit
with limited statistics. Scattering of antineutrons on nuclei made
it possible to investigate nucleon-antinucleon annihilation inside
matter, without complications due to Coulomb
interaction~\cite{Bressani:2003pv}.

In the above experiments antineutrons were produced in
proton-antiproton annihilation via charge exchange (CEX) $\pbar
p\to \nbar n$. The disadvantages of this method are obvious:
production rate is low and antineutron momentum and direction are
hard to control. Selection of antineutrons with momentum in a
specific direction results in discarding a large fraction of
antineutrons. Further higher rate experiments have been proposed,
utilizing high-intensity antiproton beams, but within this
approach one does not expect a significant improvement in the
antineutron accumulation rate.

Ground state $SU(3)$ octet hyperons ($\Lambda$, $\Sigma^{+,-}$,
$\Xi^{0,-}$) and their antiparticles have relatively long lifetime
(typical $c\tau$ of a few cm) and are essential for investigating
several important physics questions, including hyperon-nucleon
interaction and and the possible role of hyperons in neutron
stars~\cite{Haidenbauer:2013oca,Vidana:2018bdi,Tolos:2020aln}. The
relevant experimental studies started in 1960s and have lasted for
more than half a century~\cite{Engelmann:1966,SechiZorn:1969hk,
Alexander:1969cx,Eisele:1971mk,Kadyk:1971tc,Hauptman:1977hr,
Ahn:1997wa,Aoki:1998sv,Kondo:2000hn,Ahn:2005jz}, using $\pi^\pm$
or $K^\pm$ beams shot into bubble chambers or Scintillating Fiber
(SciFi) targets. The statistics of these experiments are low, with
typically a few tens to a few hundreds observed events.

In this article we propose a new source of antineutrons and
(anti)hyperons, based on a completely different technique. Due to
its large production cross section in $\EE$ and large decay
branching fractions to final states with antineutron or
(anti)hyperon, $\jpsi$ particles produced in an $\EE$ annihilation
experiment can serve as a new source of antineutrons and
(anti)hyperons. By tagging the particles recoiling against an
antineutron or (anti)hyperon in $\jpsi$ decays, the momentum and
direction of these source particles can be determined precisely,
and the interaction with the target material placed outside of a
very small radius $\EE$ beam pipe allows substantial
high-precision measurements related to particle and nuclear
physics.

We show below that antineutrons produced in $\,\jpsi\to \ppinb\,$
and (anti)hyperons produced in similar $\jpsi$ decay modes at a
tau-charm factory like BESIII~\cite{bes3} at BEPCII~\cite{bepc2},
or a super tau-charm factory like STCF~\cite{STCF} or
SCTF~\cite{SCTF} can serve as a perfect source of these particles
for many physics studies, thanks to the huge $\EE\to \jpsi$
production rate and the modern, multipurpose high-performance
detectors. We use the existing BESIII experiment as a case study,
with its accumulated 10 billions $\jpsi$-s dataset, to provide a
proof of concept of using antineutrons from $\jpsi$ decays. We
extend our discussion to $SU(3)$ octet hyperon and antihyperon
sources and to future higher-luminosity experiments under
consideration. We point out that by including in the detector
design an option for inserting a variety of specific materials as
targets for these particles, a wide range of novel high-precision
physics measurements can be carried out. These include
antinucleon-nucleon interaction~\cite{Klempt:2005pp}, OZI
violation and nonvalence $s\bar{s}$ component of the
nucleon~\cite{Ellis:1988jwa,Ellis:1994ww,Ellis:1999er},
(anti)hyperon-nucleon
interaction~\cite{Haidenbauer:2013oca,Tolos:2020aln},
(multi-strange)
hypernuclei~\cite{Gal:2016boi,Botta:2012xi,Pochodzalla:2016ncu},
light hadron spectroscopy~\cite{Rosner:2006jz,Klempt:2007cp},
including exotics and many others~\cite{Bressani:2003pv}, as well
as cross section of antineutrons with material for calibration of
Monte Carlo simulation codes for particle physics and medical
applications, such as {\sc fluka}~\cite{FLUKA} and {\sc
geant4}~\cite{GEANT}.


The $\jpsi$ was discovered in 1974~\cite{ting,richter} and has
been studied in many generations of detectors. Due to its large
partial width to $\EE$, the production cross section of $\jpsi$ in
$\EE$ annihilation is about 90~$\mu$b. However, the actual
production rate suffers from initial state radiation and the
energy spread of the $\EE$ beams~\cite{wym}. As a matter of fact,
the experimental production cross section of $\jpsi$ at BEPCII is
3500~nb with an energy spread of 0.9~MeV for the center-of-mass
(CM) energy~\cite{bes3_mm}.
The BEPCII~\cite{bepc2} is a symmetric $\EE$ collider operating in
CM energies from 2 to 5~GeV, with design luminosity of
$10^{33}~{\rm cm}^{-2}s^{-1}$ at 3.77~GeV, the peak of the
$\psi(3770)$ resonance. The luminosity at 3.097~GeV, the $\jpsi$
resonance, is about $0.47\times 10^{33}~{\rm cm}^{-2}s^{-1}$. The
energies of the $e^+$ and $e^-$ beams are known to 0.1~MeV and
they collide with a crossing angle of 22~mrad in the horizontal
plane, so the $\jpsi$ particle is basically static with a very
small boost which is known precisely. This is equivalent to a
production rate of 1600~Hz for $\jpsi$ events.

Amongst more than 300 decay modes reported, $\jpsi\to \ppinb\,$ is
the best source of antineutrons: the branching fraction, $(2.12\pm
0.09)\times 10^{-3}$, is large~\cite{pdg} and there are only two
charged tracks originating from the interaction point. They can be
selected and identified with high efficiency in a detector like
BESIII.

The BESIII detector is described in detail in Ref.~\cite{bes3}.
The cylindrical core of the detector covers 93\% of the full solid
angle and consists of a helium-based multilayer drift
chamber~(MDC) outside of a beryllium beam pipe, a plastic
scintillator time-of-flight system~(TOF), and a CsI(Tl)
electromagnetic calorimeter~(EMC), which are all enclosed in a
superconducting solenoidal magnet providing a 1.0~T magnetic
field. The solenoid is supported by an octagonal flux-return yoke
with resistive plate counter muon identification modules
interleaved with steel. The charged-particle momentum resolution
at $1~{\rm GeV}/c$ is $0.5\%$, and the $dE/dx$ resolution is $6\%$
for electrons from Bhabha scattering. The EMC measures photon
energies with a resolution of $2.5\%$ ($5\%$) at $1$~GeV in the
barrel (end cap) region. The time resolution in the TOF barrel
region is 68~ps, while that in the end cap region is 110~ps. The
end cap TOF system was upgraded in 2015 using multi-gap resistive
plate chamber technology, providing a time resolution of
60~ps~\cite{etof}. In four runs in 2009, 2012, 2018 and 2019,
BESIII has accumulated 10 billion high-quality $\jpsi$ events.
These have yielded a large number of significant published
results, based on a part of or the full data sample~\cite{bes3}.

With known four-momenta of the initial $\jpsi$ and those of the
selected proton and $\pi^-$, the antineutron can be selected by
requiring the recoiling mass of $p\pi^-$ to agree with that of an
antineutron. The momentum and direction of the antineutron can be
determined as well, with an uncertainty of a few MeV and a few
milliradians, respectively. This estimate is based on published
results on $\jpsi\to \ppinb$ from BES~\cite{pnbarpi_bes}, whose
momentum resolution is not as good as that of BESIII, and $\psp\to
\gamma\chi_{cJ}\to \gamma\ppinb$ from
BESIII~\cite{chicj_pnbarpi_bes3}, where one more photon is
involved in the final state. With $\BR(\jpsi \to p\pi^- \bar
n)\approx 2\times 10^{-3}$,\, the $\,{\sim}40$\% tagging
efficiency indicates that 8 million antineutrons can be tagged in
the 10 billion $\jpsi$ event sample.

\begin{figure*}[htbp]
\centering
\includegraphics[height=0.3\textwidth]{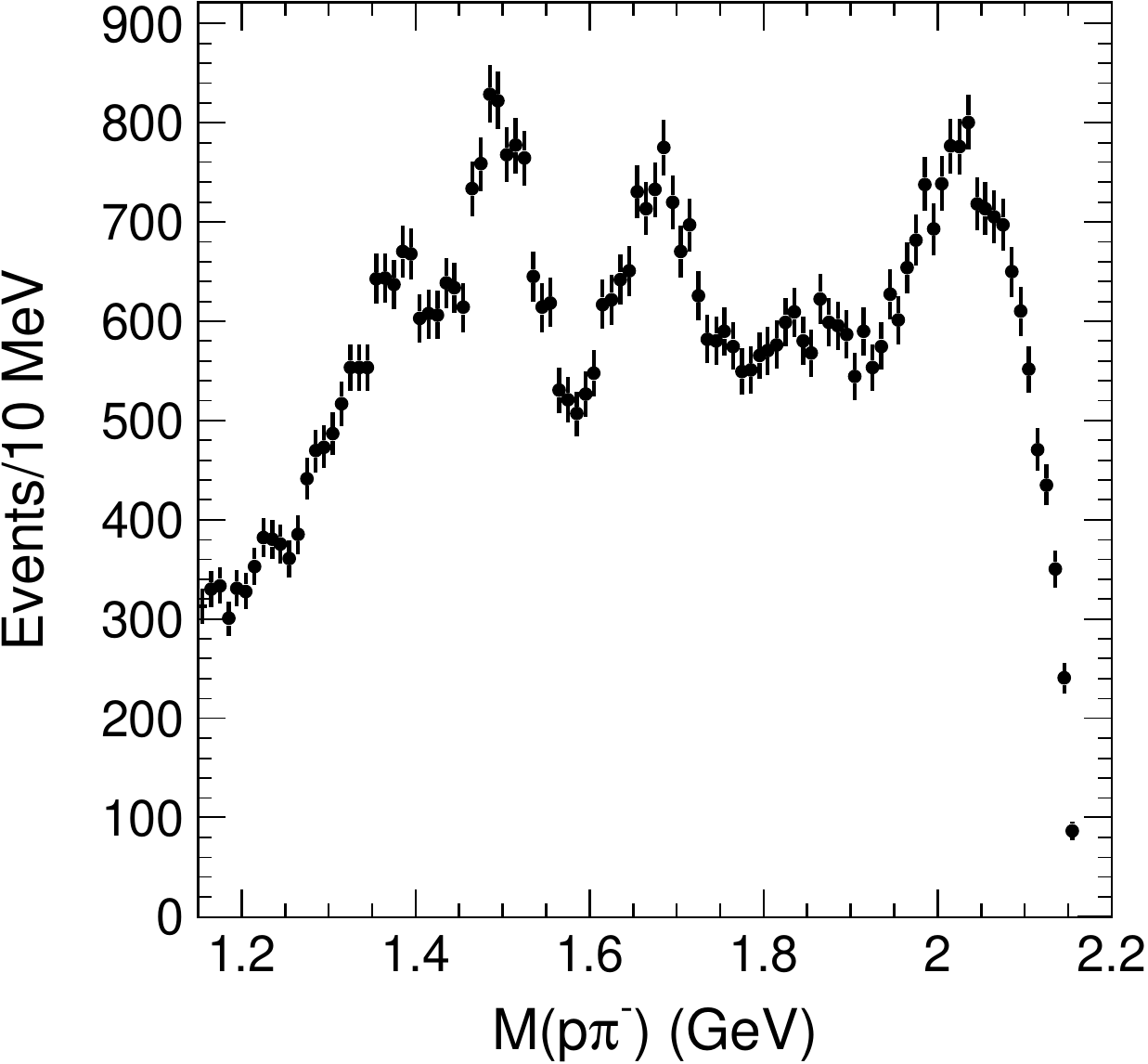}
\kern 3em
\includegraphics[height=0.3\textwidth]{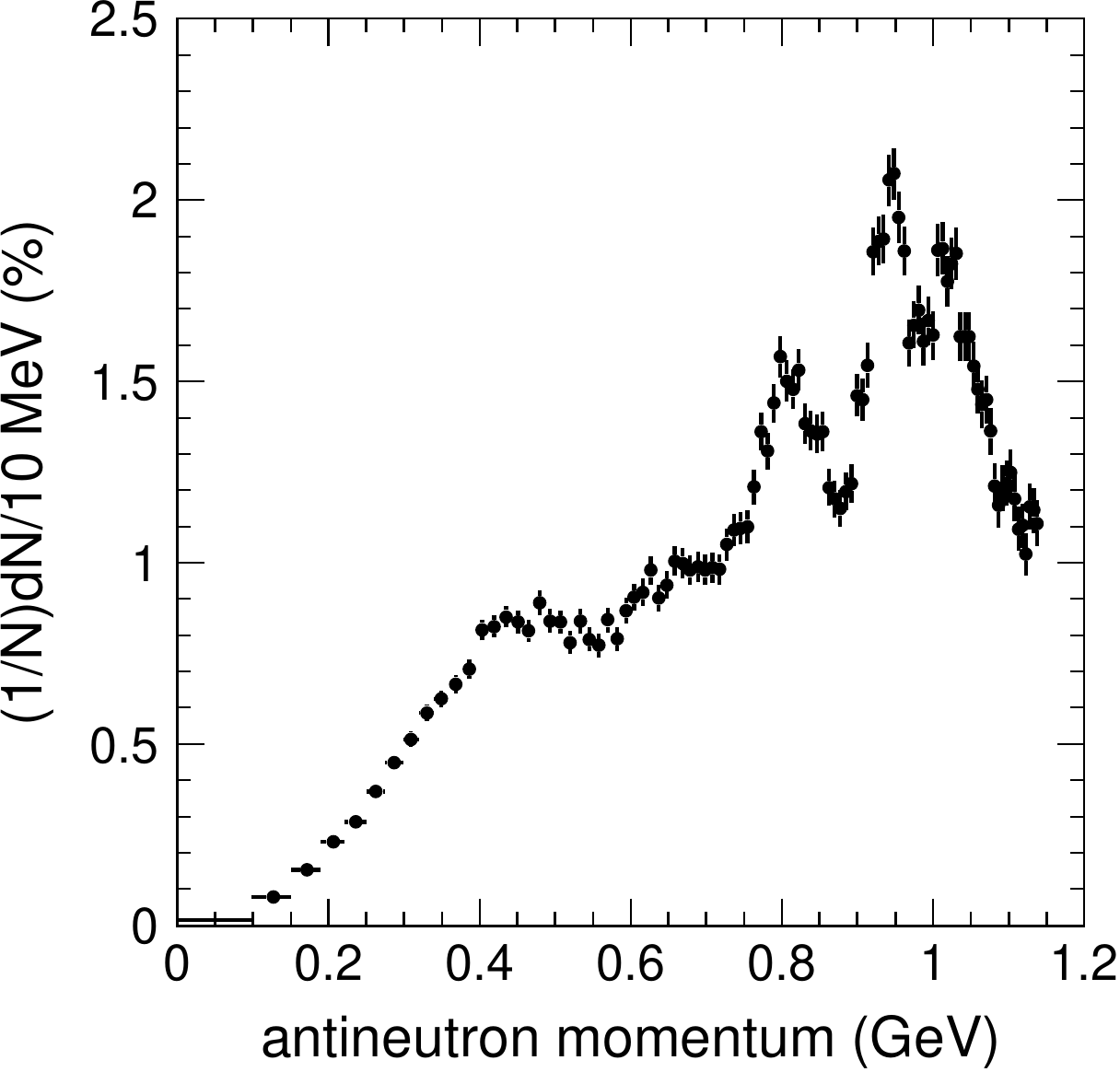}
 \kern 2em \strut
\caption{Invariant mass distribution of $p\pi^-$ from $\jpsi\to
\ppinb$ selected from BES experiment (left), and the corresponding
antineutron momentum distribution (fraction of events in
10~MeV/$c$) (right). The plots have been remade with data in
Ref.~\cite{pnbarpi_bes}.} \label{fig_mppi}
\end{figure*}

The maximum momentum of the antineutron is 1174~MeV/$c$,
corresponding to the case when the proton and $\pi^-$ fly in the
same direction, opposite to the antineutron; the minimum momentum
of the antineutron is zero, when the proton and $\pi^-$ fly
back-to-back with the same absolute value of the momentum. The
momentum spectrum of the antineutron, shown in the right panel of
Fig.~\ref{fig_mppi}, is determined by the $\jpsi\to \ppinb$ decay
dynamics. It can be obtained by converting the $p\pi^-$ invariant
mass distribution~\cite{pnbarpi_bes} (shown in the left panel of
Fig.~\ref{fig_mppi}) into antineutron momentum distribution.

Comparing this sample with that achieved by the OBELIX
experiment~\cite{Astrua:2002zg}, we find that the BESIII sample is
already a quarter of that in OBELIX and the momentum range is
wider. The momentum of the antineutron at BESIII is known with an
uncertainty of about 6--7~MeV/$c$ in the full momentum range
whereas that at OBELIX is 3-5\% of 50 to 400~MeV/$c$. In
particular, antineutrons with momentum higher than 500~MeV/$c$
(about 80\% of the tagged antineutrons) are unique in the world.
These events can be used to study many open problems involving the
physics of antineutrons, as discussed in the
review~\cite{Bressani:2003pv}.


Ground state $SU(3)$ octet hyperons ($\Lambda$, $\Sigma^{+,-}$,
$\Xi^{0,-}$) and their antiparticles can be produced copiously in
$\jpsi$ decays and they can be tagged similarly to antineutrons.
Since their $c\tau$ is on the order of a few cm, comparable to
inner radius of the beam pipe (see below), only some make it to
the beam pipe and/or the inner tube of the MDC, but the surviving
fraction is significant. This makes the study of hyperon and
antihyperon interactions with nuclei feasible.
Table~\ref{tab:hyper} lists the main $\jpsi$ decay modes relevant
for hyperon/antihyperon production, together with their $c\tau$,
to be compared with the geometry of BESIII beam pipe and of MDC
inner tube. A rough estimate of the available tagged particles in
the 10 billion $\jpsi$ event sample can be made according to the
tag efficiency (${\sim}20\%\div40\%$, depending on multiplicity of
final state particles), the $\beta\gamma$ of the produced hyperons
and their lifetimes. The numbers of particles reaching the beam
pipe (as target material at BESIII) are listed in
Table~\ref{tab:hyper}.

\begin{table*}[htbp]
\def\mystrut{\vrule width 0pt height 2.5ex depth 1.5ex}
\def\upstrut{\vrule width 0pt height 2.5ex}
\caption{Hyperon and antihyperon production in the 10 billion
$\jpsi$ or 3 billion $\psp$ event data sample at BESIII. The yield of
hyperons is the same as that of antihyperons, since particles and
antiparticles are produced with the same rates in $\jpsi$ (or
$\psp$) decays via strong or electromagnetic interactions. $p_{\rm
max}$ is the maximum momentum of the antihyperon, $n^Y_{\rm BP}$
is the number of tagged antihyperons reaching the beam pipe;
``---'' means not available.}
    \label{tab:hyper}
    \centering
    \begin{tabular}{cclccc}
    \hline\hline
    \mystrut
      Antihyperon & ~~$c\tau$ (cm)~~ &  decay mode
                                  & Branching Fraction~($\times 10^{-3}$)
                                  & ~~$p_{\rm max}$ (MeV/$c$)~~ & $n^Y_{\rm BP}$ ($\times 10^5$)\\\hline
  $\bar{\Lambda}$ & 7.89  & $\jpsi\to \Lambda\bar{\Lambda}$ &  1.89  &  1074 & 26 \upstrut\\
                  &       & $\jpsi\to pK^-\bar{\Lambda}$    &  0.87  &   876 & 9\\\hline
  $\bar{\Sigma}^-$ & 2.40 & $\jpsi\to \Sigma^+\bar{\Sigma}^-$ & 1.50 &   992 & 4 \upstrut\\
                   &      & $\jpsi\to \Lambda \pi^+\bar{\Sigma}^-$   &  0.83   &  950 & 1 \\\hline
  $\bar{\Sigma}^+$ & 4.43 & $\jpsi\to \Lambda \pi^-\bar{\Sigma}^+$   &  --- &  945  & --- \upstrut \\\hline
   $\bar{\Xi}^0$ & 8.71 & $\jpsi\to \Xi^0\bar{\Xi}^0$ &  1.17   &  818 & 7 \upstrut \\
                 &      & $\jpsi\to \Xi^-\pi^+\bar{\Xi}^0$ &  ---  &  685 & ---\\\hline
   $\bar{\Xi}^+$ & 4.91 & $\jpsi\to \Xi^-\bar{\Xi}^+$ &  0.97   &  807 & 3 \upstrut \\
                 &      & $\jpsi\to \Xi^0\pi^-\bar{\Xi}^+$ &  ---  &  686 & ---\\\hline
   $\bar{\Omega}^+$ & 2.46 & $\psp\to \Omega^-\bar{\Omega}^+$ &  0.05   &  774 & 0.05 \upstrut \\
                    &      & $\psp\to K^-\Xi^0\bar{\Omega}^+$ &  ---  &  606 & --- \\
      \hline\hline
    \end{tabular}
\end{table*}

The $\Omega^-$ and its antiparticle $\bar{\Omega}^+$ can not be
produced in $\jpsi$ decays, due to their high mass, but they are
accessible through $\psp$ decays. As $\BR[\psp\to
\Omega^-\bar{\Omega}^+]=5.2\times 10^{-5}$, they can be studied
with the BESIII $\psp$ data sample~\cite{bes3}. There could also
be other $\psp$ decay modes with $\Omega^-$ or $\bar{\Omega}^+$ in
the final state, such as $\psp\to K^-\Xi^0\bar{\Omega}^+ + c.c.$
These, too, are listed in Table~\ref{tab:hyper}, together with the
number of $\Omega$s reaching the beam pipe in the 3 billion $\psp$
event data sample~\cite{bes3}. Of course the $\psp$ data sample
will be able to supply more decay modes for all these hyperons and
antihyperons, including the decays from the secondary charmonium
states [$\chi_{cJ}$~($J=0$, $1$, $2$), $\jpsi$] produced in $\psp$
radiative and hadronic transitions~\cite{note}. These new decay
modes will broaden the momentum range of the source particles.


The BESIII $\jpsi$ data sample has been collected already, so we
do not have an opportunity to gather data with a custom-made
target. However, the detector material close to the interaction
point in the inner detector serves as an effective target,
allowing us to carry out a substantial study with the collected
antineutrons.

The BESIII~\cite{bes3} beam pipe is 1000~mm long with an inner
diameter of 63~mm and an outer diameter of 114~mm. The central
part of the beam pipe is 296~mm long with two layers of beryllium.
The inner wall that maintains the ultra high vacuum in the beam
pipe is 0.8~mm thick and the outer beryllium wall is 0.6~mm thick.
A 0.8~mm channel between the two walls is used for circulating the
cooling fluid (high-purity mineral oil). Outside of the beam pipe
is the MDC. The MDC consists of an outer chamber and an inner
chamber, which are joined together at the end plates, sharing a
common gas volume. The inner tube of the MDC is made of 1.2~mm
thick carbon fiber with a radius of 59.2~mm that also provides
some mechanical strength. The inner chamber of the MDC, together
with the inner tube can be replaced in case it is damaged by
radiation. This design even allows us put some target material
into the current BESIII detector if we want to do antineutron
physics in the near future, as will be discussed below.

\begin{figure}[htbp]
\centering
\includegraphics[width=0.55\textwidth]{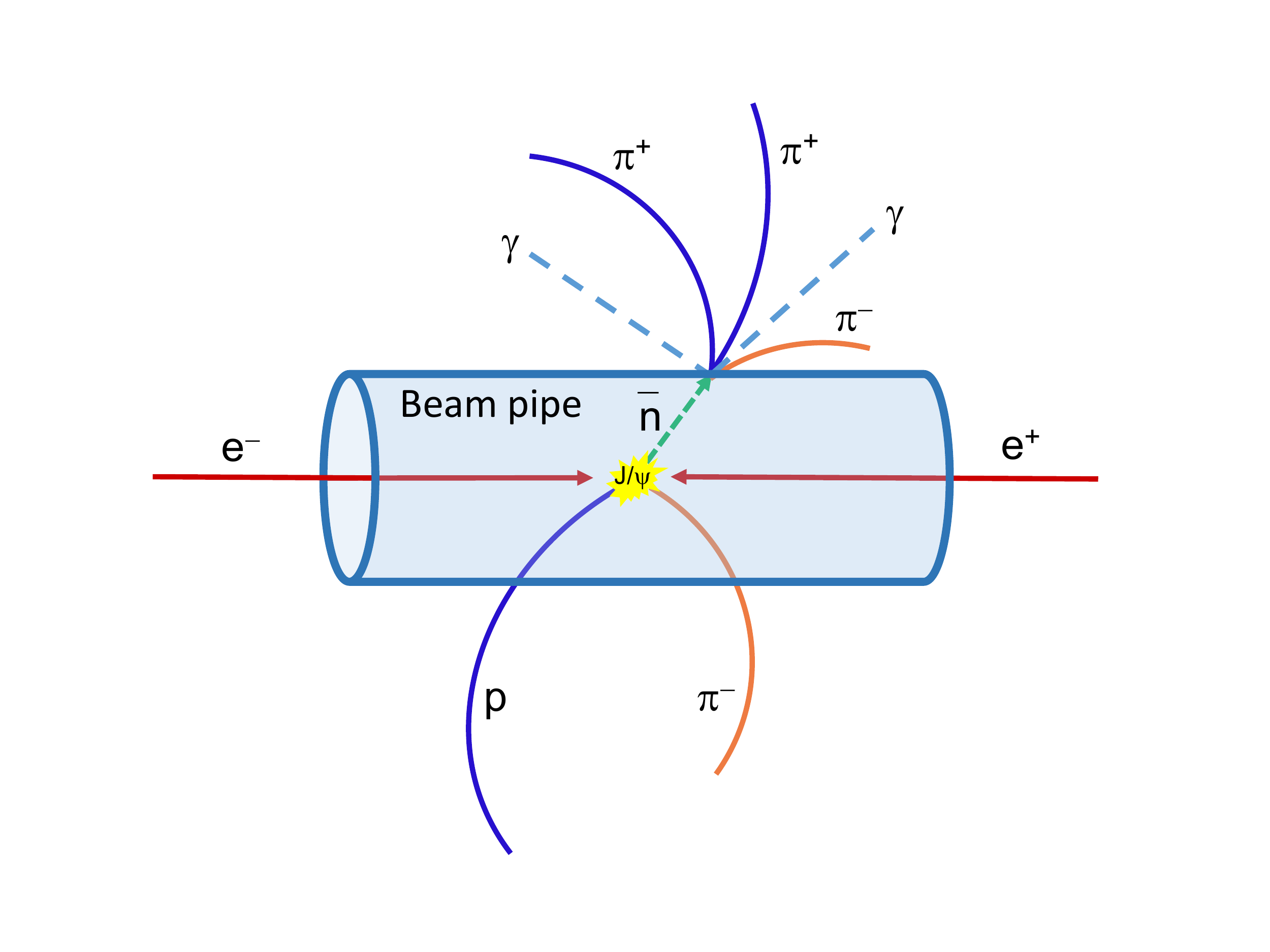}
\caption{Schematic diagram of $\EE\to \jpsi\to \ppinb$, followed
by $\nbar$ interaction with a proton in the beam pipe material,
$\nbar p \to \pi^+\pi^+\pi^-\pi^0$, $\pi^0\to \gamma\gamma$.}
\label{fig_schem}
\end{figure}

The materials in the beam pipe and the inner tube of the MDC can
be treated as targets of beryllium and carbon. Since all $\bar n$
interactions with these materials occur in the central part of the
detector, all the final state particles can be detected by the
full detector, just like those produced in the primary $\jpsi$
decays, in terms of tracking, particle identification, interaction
vertex determination, photon reconstruction and so on. A schematic
diagram of $\EE\to \jpsi\to \ppinb$, followed by $\nbar$
interaction with a proton in the beam pipe material is shown in
Fig.~\ref{fig_schem}. The charged pions and the photons from
neutral pion decays are shown for a possible process $\nbar p \to
\pi^+\pi^+\pi^-\pi^0$, $\pi^0\to \gamma\gamma$.

The good momentum resolution of the BESIII detector makes it
possible to pinpoint the interaction vertex with a resolution of
about 1~mm~\cite{vertex}, so as to determine whether an
antineutron interacted with beryllium or carbon, but the
resolution is not sufficient to separate interaction with inner or
outer wall of the beam pipe, nor with the mineral oil between the
two layers of beryllium. The background from direct $\jpsi$ decays
is low, since all the final state particles are reconstructed and
the total energy of the event is larger than the background $\EE$
CM energy. The background from particle interaction with target
material, but with final states different from the one under study
can be suppressed thanks to the good particle identification and
good momentum/energy resolution of the detector.

One can use the materials' budget to estimate the fraction of
antineutrons interacting with specific components of the inner
detector. Assuming an average of 100~mb cross section per $\nbp$
or $\nbn$ collision, we expect 1--2\% of the tagged antineutrons
to interact with beryllium and another 1--2\% with carbon fiber
targets. These are equivalent to about 100,000 events each for
$\nbar+{\rm Be}$ and $\nbar+{\rm C}$ interactions, respectively.


Accelerators at tau-charm energy region with luminosity 100 times
higher than current one are being proposed, focusing on a variety
of physics associated with strong and electroweak interactions.
Both STCF~\cite{STCF} and SCTF~\cite{SCTF} operate in a way
similar to BEPCII. The detectors will be even better than BESIII
in term of tracking, particle identification, and photon
detection. Experiments at these new facilities will be able to
accumulate one trillion ($10^{12}$) $\jpsi$ events in one year's
running time, i.e. 100 times more than the total dataset
accumulated by BESIII.

If one were to just simply rescale the numbers in previous
sections accordingly, one would immediately see that such a
machine provides a superb opportunity for physics with baryon and
antibaryon sources discussed above. In fact, the case is even more
striking, since several additional features in the accelerator and
the detector design will result in significant improvements in the
quality of both the sources and the targets.

In the current BEPCII accelerator the energy spread at the $\jpsi$
peak is 0.9~MeV. Since the natural width of $\jpsi$ is only about
90~keV, the peak cross section for $\jpsi$ production will be
significantly increased once the beam energy spread is further
reduced. It is expected that use of the monochromator scheme will
significantly reduce the energy spread and increase the peak cross
section, by a factor between a few times and orders of
magnitude~\cite{mono1,mono2,mono3}.

The range of physics topics investigated with these baryons and
antibaryons and be expanded through use of custom-made targets can
be designed and placed inside the detector, just outside of the
beam pipe. The radius of the beam pipe can be made smaller than
that at BEPCII, allowing the targets be located closer to the
interaction point, enlarging acceptance of the target material and
reducing the fraction of hyperon decays. In principle one can
design a barrel with several different types of materials as
targets, allowing simultaneous study of different interactions.
This will be extremely efficient, since all studies share the same
environment and systematic effects can be investigated once for
all. Of course, removable targets will be helpful for even more
studies. This does not involve technical difficulties, as the
current BESIII~\cite{bes3} design has a replaceable inner drift
chamber, and the inner tube of the chamber has been used as a
target in the data taking.

In order to produce baryon and antibaryon particles with higher
momenta, one can use asymmetric $e^+e^-$ beams which will produce
$\jpsi$ in motion. Alternatively, one can utilize similar decay
modes of $\psp$, or both. One may also broaden the momentum
spectra of baryons produced in $\jpsi$ decays, by utilizing
$\jpsi$ boosted through large crossing angle
collision~\cite{mono3}.

Detector design can be optimized to enhance the detection of the
final states specific to the reactions discussed in the present
work. The current detector design is optimized for measurements of
light particles such as $e$, $\mu$, $\pi$, $K$, $p$, and photon. A
special-purpose subdetector can be devised for identification of
particles, like deuteron, triton, and even heavier nuclei.


In summary, we demonstrate that $\jpsi$ produced in
high-luminosity $\EE$ annihilation can provide large numbers of
baryon and antibaryon particles as sources for novel nuclear and
particle physics studies. These sources include all the long-lived
baryons and antibaryons, especially antineutron, hyperons
$\Lambda$, $\Sigma^{+,-}$, $\Xi^{0,-}$, and their antiparticles,
as well as smaller, but still large fluxes of $\Omega^-$ and
$\bar\Omega^+$. By placing specific custom-made targets in a
BESIII-like detector, one can perform a rich variety of
experiments in nuclear and particle physics. Besides those
discussed in the paper or listed in Table~\ref{tab:hyper}, other
decay modes with antineutron or hyperons can also be used as
sources of these particles. Although the tagging efficiency may be
low in multi-particle final states, more low momentum source
particles can be produced in these modes.

With the existing 10 billion $\jpsi$ event data sample accumulated
at BESIII, many studies of the interactions of these long-lived
baryons with the material in the beam pipe and the inner tube of
the MDC detector can already be performed. With minor
modifications of the  accelerator and detector designed for
STCF~\cite{STCF} and SCTF~\cite{SCTF}, the resulting $\jpsi$
sample can serve as a superb source of all kinds of long-lived
baryons and antibaryons, opening a new era for novel
high-precision nuclear and particle physics studies.

Traditional setups need to produce many different kinds of beams
for different dedicated experiments, and need to share accelerator
time among them. This requires large resources in terms of
manpower and funding, impeding such experiments. In contrast, the
approach proposed here will allow experiments with different beams
at the same time, requiring no additional infrastructure and
minimal further investments.


This work is supported in part by National Key Research and
Development Program of China under Contract No.~2020YFA0406300,
National Natural Science Foundation of China (NSFC) under contract
Nos. 11961141012, 11835012, and 11521505; and the CAS Center for
Excellence in Particle Physics (CCEPP), as well as NSFC-ISF grant
No.\ 3423/19. C.Z.Y. thanks Sicheng Yuan for stimulating
conversations about antineutron-material interactions, and M.K.
thanks Avraham Gal, Simon Eydelman and Jon Rosner for helpful
discussions.



\begin{thebibliography}{000}
\bibitem{note} It should be noted that particles other than those
which are the focus of this paper are also produced copiously in
$\jpsi$ and $\psp$ decays. With similar methods, $\pi^\pm$,
$K^\pm$, $K^0/\bar{K}^0$, $p/\bar{p}$, neutron, photon, and other
particles can be also be tagged (for $K_L$
cf.~\cite{Amaryan:2020xhw}). Although the corresponding samples
may not be large compared with those from dedicated experiments,
they are produced for free, their momentum is known with high
precision and they can be used for physics studies with the same
targets and detector. In a super $\jpsi$ factory the corresponding
yields will be higher by two orders of magnitude or more, so in
certain cases the samples may become competitive with existing
data from dedicated experiments.
\end{thebibliography}
\end{document}